\documentclass[showpacs,preprint]{revtex4}
\usepackage{amssymb,amsmath}

\newcommand{\eqn}[1]{(\ref{#1})}
\def\appendix#1{\addtocounter{section}{1}\setcounter{equation}{0}
\renewcommand{\thesection}{\Alph{section}}
\section*{
\thesection\protect\indent \parbox[t]{11.715cm} {#1}}
\addcontentsline{toc}{section}{Appendix\thesection\ \ \ #1} }
\newcommand{\zed}{{\mathbb Z}} 
\newcommand{\real}{{\mathbb R}} 

\def\nn{\nonumber}

\newcommand{\be}{\begin{equation}}
\newcommand{\ee}{\end{equation}}
\newcommand{\beq}{\begin{equation}}
\newcommand{\eeq}{\end{equation}}
\newcommand{\bea}{\begin{eqnarray}}
\newcommand{\eea}{\end{eqnarray}}

\newcommand{\del}{\partial}

\begin{document}

\preprint{DSF--49--2005} \preprint{hep--th/0601056}
\hfill{ }\\

\title{Twisted Conformal Symmetry in Noncommutative Two-Dimensional
                \\ Quantum Field Theory}

\author{Fedele Lizzi$^*$, Sachindeo Vaidya$^\dagger$ and Patrizia Vitale$^*$}
\affiliation{$^*$Dipartimento di Scienze Fisiche,\\Universit\`a di Napoli
  Federico II and INFN Sezione di Napoli, \\Via Cintia, 80126,
  Napoli, ITALY. \\~\\
$^\dagger$Centre for High Energy Physics,~Indian Institute of Science,
  \\Bangalore, INDIA 560012.
}

\date{\today}

\begin{abstract}
By twisting the commutation relations between creation and
annihilation operators, we show that quantum conformal invariance can
be implemented in the 2-d Moyal plane. This is an explicit realization
of an infinite dimensional symmetry as a quantum algebra.
\end{abstract}

\maketitle

{\it Dedicated to Rafael Sorkin on the occasion of his 60th birthday.}

Conformal theories on the two-dimensional plane \cite{BPZ} play an
important role in several aspects of modern physics, from string
theory to applications to condensed matter. In this article, we will
investigate the fate of this infinite dimensional symmetry when the
plane is deformed to become the simplest \emph{noncommutative
geometry}. We will consider a plane in which the coordinates do not
commute $[x_0,x_1]_\star=i\theta$, where $\theta$ is a constant with
the dimensions of the square of length (in the units $c=1$), and the
meaning of the $\star$ subscript will be clear in the following. We
will demonstrate that it is possible to construct a \emph{quantum}
field theory that is \emph{quantum} both in the sense that fields
are operators on a Fock space, and that the symmetry is realized as
a quantum algebra. This provides an explicit example of the common
believe that quantum algebras describe the symmetries of
noncommutative geometries. An alternative approach to conformal
symmetry on the 2-d Moyal plane which has points of overlap with
this work is due to A.~P.~Balachandran, A.~Marques and
P.~Teotonio-Sobrinho (to appear).

It is sometimes argued that the presence of an intrinsic length scale
is incompatible with conformal invariance. This is no more correct
than saying that the existence of a fundamental constant with
dimensions of angular momentum necessarily destroys rotational
invariance.The point is the manner in which the symmetry is
implemented on the space of interest.

Consider fields on the noncommutative plane ${\real}^2_\theta$.
Functions on this space are multiplied with the Moyal star product
\begin{equation}
(f \star g)(x) = e^{\frac{i}{2}\theta (\partial_{x_0} \partial_{y_1} -
    \partial_{x_1} \partial_{y_0})} f(x) \cdot g(y)|_{x=y}
\end{equation}
For our purposes it is useful to see this product written as an ordinary
product $m_0$ of a ``twisted'' tensor product:
\begin{equation}
(f \star g)(x) = m_0 [{\cal F}^{-1} f \otimes g] \equiv
    m_\theta [f \otimes g] \label{moyal}
\end{equation}
where $m_0(f\otimes g)=f\cdot g$ is the ordinary product and
\begin{equation}
{\cal F} = e^{-\frac{i}{2} \theta^{\mu\nu} \partial_{x_\mu}
\otimes
\partial_{y_\nu}} = e^{-\frac{i}{2} \theta(\partial_{x_0}
\otimes
\partial_{y_1}- \partial_{x_1} \otimes \partial_{y_0})}
\label{twist}
\end{equation}
is the {\it twist}.

Our starting point is the observation
in~\cite{Wess,CKNT,ChaichianPresnajderTureanu} that when a product
is a twist deformation of the commutative product, infinitesimal
symmetries that act on the space are implemented by deforming the
coproduct by the same twist. In~\cite{oeckl, gonera} a global
approach is undertaken, where the infinitesimal symmetries, together
with their deformed coproduct, are seen to descend from a
deformation of the product on the group manifold. The original study
concerned the Poincar\'e symmetry on ${\real}^{1,3}_\theta$, and
there has been subsequent work for conformal
symmetries~\cite{Matlock} in dimensions greater than 2, as well as
diffeomorphisms~\cite{ABDMSW} and gauge transformations~\cite{vas}.

To understand how infinitesimal generators of a given Lie algebra
$ \mathcal{G}$ act on the noncommutative plane, let us have a new
look at their action in the commutative case.   Given an
infinitesimal generator $X$, we have
\be X (f\cdot g)= X f \cdot g + f \cdot X g
\ee
which we can rewrite as
\be X (f\cdot g)= m_0[
(\mathbb{I}\otimes X + X\otimes  \mathbb{I})(f\otimes g)]\ee where
\be \mathbb{I}\otimes X + X\otimes  \mathbb{I} \equiv
\Delta_0
\ee is the standard coproduct in the enveloping algebra
$ U(\mathcal{G})$. It captures the ordinary Leibniz rule. However
with the $\star$ product the rule is violated, infinitesimal
generators are not anymore derivations of the new algebra:
\be
X (f\star g )\ne Xf\star g + f \star X g
\ee
This means that the coproduct in the enveloping algebra has to be
changed, hence $ U(\mathcal{G})\rightarrow U_\theta (\mathcal{G})$.
Indeed we have
\bea
     X (f\star g) &=& X \cdot m_\theta (f\otimes g)
      = X\cdot m_0\big(\mathcal{F} ^{-1}\cdot(f\otimes g)\big) \nonumber\\
       & =& m_0\big(\Delta(X)\mathcal{F}^{-1}\cdot(f\otimes
       g)\big)\nn\\
        &=& m_0\big(\mathcal{F}^{-1}\Delta_\theta(X)\cdot(f\otimes g)\big)  \nonumber\\
       & =& m_\theta\big(\Delta_\theta(X)\cdot(f\otimes g)\big)
          \label{infsym}
\eea
where the new coproduct is
\be
\Delta_\theta(X)=\mathcal{F}^{-1}\Delta_0 (X) \mathcal{F}.
\label{twistcoprod}
\ee
The  Lie algebra itself doesn't have to be changed. Namely, the
action on single functions is {\it undeformed}. Notice, as a
technical remark, that the deformed enveloping algebra  $U_\theta
(\mathcal{G})$, is a noncommutative, noncocommutative Hopf algebra,
i.e.\ a \emph{Quantum Group}.

Therefore, symmetries of noncommutative theories should be
investigated in the form \eqn{infsym}. A noncommutative theory will
 have a quantum symmetry if it is left invariant under the infinitesimal action
  \eqn{infsym}.
The authors above cited dealt with the classical symmetries of the
Moyal space, while the extension of twisted symmetry to the quantum
domain was done in~\cite{bmpv}. To implement quantum symmetries we
will take the point of view that whenever the tensor product of two
fields appears, one always applies the twist. This means that all
commutator among fields will automatically be $\star$-commutators,
the representations will be combined with deformed coproducts, and
that two-point correlation functions are calculated twisting the
product of fields.

We want to investigate the fate of conformal symmetry in two
space-time dimensions, when space-time is made noncommutative. To
this,  let us consider the two dimensional Minkowski plane in
light cone coordinates. A treatment using complex coordinates is
probably possible, and definitely interesting, but we will not do
it here. We will use the convention in which the spacetime metric
$\eta_{\mu \nu} ={\rm diag}(1, -1)$. Light cone coordinates are
defined as $x^\pm = x^0 \pm x^1$. Using $\eta_{AB} dx^A dx^B =
\eta_{\mu \nu}dx^\mu dx^\nu$, and $\eta^{AB}=(\eta_{AB})^{-1}$
(here $A, B$ label lightcone indices + and -) we have
\begin{eqnarray}
\eta_{++} = \eta_{--} &=0=& \eta^{++} = \eta^{--}, \nonumber \\
\eta_{+-} = \eta_{-+} &= \frac{1}{2}=&(\eta^{+-})^{-1} =
(\eta^{-+})^{-1}.
\end{eqnarray}
{}From $x_A = \eta_{AB} x^B$ we also have the rule $ x_\pm =
{x^\mp}/{2}\ .$

Let us briefly recall the situation in the commutative case. In two
dimensions the infinite dimensional conformal algebra is generated by
vector fields $u^\mu (x)\del_\mu$ that satisfy
\begin{equation}
\partial_\mu u_\nu + \partial_\nu u_\mu - \eta_{\mu \nu}
\partial^\alpha u_\alpha =0.
\end{equation}
In light cone coordinates, this is $\partial_+ u_+ =0 = \partial_-
u_-$, which implies that
\begin{eqnarray}
u_+ = u_+ (x^+) &\Leftrightarrow& u^- = u^- (x_-)
\quad {\rm and}  \nonumber \\
u_- = u_- (x^-) &\Leftrightarrow& u^+ = u^+ (x_+)
\end{eqnarray}
The $+$ and $-$ sectors are independent under conformal
transformations. Considering $u_{\pm}$ as being generated by
$-(x^{\pm})^{n+1}$, we have that conformal transformations are in turn
generated by the vector fields
\begin{equation}
\ell^\pm_n = -(x^\pm)^{n+1}\partial_\pm , \quad {\rm with} {\quad} n
\in {\zed}.
\end{equation}
They form two copies of
the classical Virasoro (Witt) algebra
\begin{eqnarray}
[\ell^\pm_n,\ell^\pm_m]&=&(m-n) \ell^\pm_{m+n}\nonumber\\
{}[\ell^+_n,\ell^-_m]&=&0
\label{VirasoroWitt}
\end{eqnarray}
On the noncommutative plane we still implement conformal
transformations via the $\ell_n$'s.
The twist~\eqn{twist}
however causes a change in the coalgebra structure.
Using~\eqn{twistcoprod} we have
\begin{widetext}
\bea
\Delta_\theta (\ell^+_n) &=& ({\bf 1} \otimes x^+ -
\theta\del_-\otimes {\bf 1})^{n+1}(-{\bf 1}\otimes\del_+)\,+\, (
x^+\otimes{\bf 1} - {\bf
1}\otimes\theta\del_-)^{n+1}(-\del_+\otimes{\bf 1}) \nonumber\\
\Delta_\theta (\ell^-_n) &=& (-{\bf 1}\otimes\del_-)({\bf 1}
\otimes x^- - \theta\del_+\otimes {\bf 1})^{n+1} \,+\,
(-\del_-\otimes{\bf 1})( x^-\otimes{\bf 1} - {\bf
1}\otimes\theta\del_+)^{n+1}. \label{twistcoprodell}
\eea
\end{widetext}
(These relations have also been independently derived by
S.~Kurkcuoglu~\cite{sk}.) The fact that the commutation relations of the Lie
algebra are unchanged, and that $\theta$ is only present in the
coproduct means that the effect of the twist is only relevant when
fields are combined, for example in correlation functions.

The above basis, while convenient for deriving the commutation
relations between generators, is inappropriate to study the
invariance of  quantum theories because it is not normalizable. Our
derivation of the twisted version of the quantum Virasoro algebra
will be in a different guise. To this end, consider the simplest two
dimensional conformally invariant theory, a scalar massless field
theory. On the noncommutative plane we have to use the star product:
\begin{equation}
S=\int d^2x ~ \del_+\varphi \star \del_-\varphi \ .
\end{equation}
The classical noncommutative action can be seen to be
twist-conformally invariant, along the lines of \cite{CKNT},
provided one uses the appropriate twist of the generators
\eqn{twistcoprodell}.  The classical solutions are fields split in
``left'' and `right'' movers $
\varphi=\varphi_+(x^+)+\varphi_-(x^-)$. Although this theory is
free (and the star product would drop from the integral), it is
still necessary to check its invariance under the action of the
conformal coalgebra. An analysis of other theories and the
behaviour of the S-matrix, including the UV/IR mixing (or lack of
it) will be presented in a longer paper in preparation.

In the quantum theory the fields are operators with mode expansion
\be  \phi(x^0,x^1) = \int_{-\infty}^\infty \frac{dk^1}{4\pi
k_0}\left( a(k) e^{-ik^\mu x_\mu} + a^\dag (k)e^{ik^\mu x_\mu}
\right). \ee Using $k^0=k_0=|k^1|$  this can be rewritten as
\begin{widetext}
\be  \phi(x^+,x^-) = \int_{-\infty}^0 \frac{dk^1}{4\pi |k^1|}\left(
a(k) e^{- i|k^1| x^+} + a^\dag (k)e^{i|k^1| x^+} \right) +
\int_{0}^{\infty} \frac{dk^1}{4\pi |k^1|}\left( a(k) e^{-ik^1 x^-} +
a^\dag (k)e^{ik^1 x^-}\right) \label{phi}. \ee
\end{widetext}
This in turn may be rewritten as
\be  \phi(x^+,x^-) =\int_{-\infty}^\infty  \frac{dk^1}{4\pi |k^1|}\left(
a_-(k) e^{- ik^1 x^+} + a_+(k) e^{- i k^1 x^-}\right)
\ee
 where we have introduced the symbol
$a_\sigma(k), ~\sigma=+,-, ~k\in(-\infty,\infty)$ related to the two
sets of oscillators appearing in \eqn{phi} (left and right movers)
by $a_\sigma(k)=a(\sigma k),~ a_\sigma(-k)=a^\dag(-\sigma k),~
k\in(0,\infty)$ .

In the standard case the commutation relations for the creation and
annihilation operators are
\begin{equation}
[a_\sigma(p),a_{\sigma'}(q)]= 2 p\delta(p+q)\delta_{\sigma\sigma'}.
\label{standcomma}
\end{equation}
Then the quantum currents $  J^+(x)=\del^+ \phi,~~   J^-(x)= \del^-
\phi$ generate two commuting U(1) Ka\c{c}-Moody algebras with
central extension.  Quantum conformal invariance is proved showing
that the components of the quantum stress-energy tensor
\begin{eqnarray}
\Theta_{\pm\pm} &=& \frac{1}{4}(\Theta_{00}\pm2
\Theta_{01}+\Theta_{11}),\nonumber \\
\Theta_{+-} &=& \frac{1}{4}(\Theta_{00} - \Theta_{11}) =
\Theta_{-+}
\end{eqnarray}
generate the conformal algebra. Tracelessness and conservation
($\partial^\mu \Theta_{\mu \nu}=0$) imply $ \Theta_{\pm\mp} =0$
and $\partial^\pm \Theta_{\pm\pm} = 0$. Hence $\Theta_{\pm\pm} (=
\frac{\Theta^{\mp\mp}}{4})$ is a function of $x^\mp$ only, as
in the standard case. Classically, $\Theta^{++}(x) = J^+(x)
J^+(x)$. But as is well-known, the quantum stress-energy tensor is
the {\it normal-ordered} product
\be
\Theta^{\pm\pm}(x)= \gamma  :J^\pm(x)
J^\pm(x):\ , \label{defLquant}
\ee
where $\gamma$ is a real number which gets fixed in the quantum
theory, and normal ordering is defined as
\bea
:a_\sigma(p)a_\sigma(q):&=&a_\sigma(p)a_\sigma(q)\ \mbox{if}\  p<q\nonumber\\
:a_\sigma(p)a_\sigma(q):&=&a_\sigma(q)a_\sigma(p)\ \mbox{if}\  p\geq
q .
\eea
Therefore the existence of Ka\c{c}-Moody quantum current algebras
is a sufficient condition to ensure conformal invariance at the
quantum level. In the noncommutative case things do not carry
through unchanged  and some crucial adjustments have to be
performed.

The Ka\c{c}-Moody currents are
\bea
J^-  &=& i \int_{-\infty}^0 \frac{dk^1}{2\pi}\left(   a^\dag (k)e^{-ik^1 x^+}
- a(k) e^{ ik^1 x^+}\right) \nonumber\\
J^+ &=& i \int_0^{\infty} \frac{dk^1}{2\pi}\left(   a^\dag (k)e^{ik^1
  x^-} - a(k) e^{- ik^1 x^-} \right) \label{KC}
\eea
where the oscillators in $J^-, J^+$ are different, being connected
to negative and positive frequencies respectively. We omit the
spatial index for the momenta from now on.

These currents, \emph{with the standard commutation relations}
\eqn{standcomma} do not give rise to the Ka\c{c}-Moody
algebras. In fact there is no reason to expect the same
commutation relations for the noncommutative plane.

To this end we pursue the following strategy for the quantum field
theory on the Moyal plane. We require that the noncommutative
Ka\c{c}-Moody relations be the same as the usual ones, but with
$\star$-commutators instead of ordinary ones:
\begin{eqnarray}
\left[J^{\pm}(x),J^{\pm}(y)\right]_\star &=&
-\frac{i}{\pi}\del_{\mp}
\delta(x^\mp - y^\mp),\nn \\
\left[J^+(x),J^-(y)\right]_\star &=& 0.
\label{starKM}
\end{eqnarray}
This will yield a deformation in the commutation rules of
creation and annihilation operators.

Let us consider the current commutators $ ~\left[J^\pm(x),
J^\pm(y)\right]_\star,~~
\left[J^+(x),J^-(y)\right]_\star ~$.
 We combine fields at different points using the twist, so that,
(with a slight abuse of notation)
\begin{widetext}
\begin{equation}
f(x)\star f(y)= \left[{\cal F} (f\otimes
g)\right](x,y)=e^{\frac{i}{2}\theta^{\mu
    \nu}\partial_{x_\mu} \partial_{y_\nu}} f(x) f(y).
\end{equation}
Our first observation is that the deformation in the product does not affect
the  commutators $ ~\left[J^\pm(x), J^\pm(y)\right]_\star$  because in
light-cone coordinates we have
\begin{equation}
e^{\frac{i}{2}\del_x\wedge\del_y} e^{\mp ikx^\pm } e^{\mp ipy^\pm}
= e^{i\theta(\del_{x^-} \del_{y^+} -\del_{x^+} \del_{y^-})} e^{\mp
ikx^\pm } e^{\mp ipy^\pm}= e^{\mp ikx^\pm } e^{\mp ipy^\pm} \
\end{equation}
\end{widetext}
with $\del_x\wedge\del_y = \theta^{\mu \nu}\partial_{x_\mu}
\partial_{y_\nu}$. Hence in each chiral sector the symmetry is
unchanged. This is to be expected since the $\star$-product between
two functions of $x^+$ (or $x^-$) alone is the same as the usual
product.  Likewise the coproduct\eqn{twistcoprodell} when acting on
pairs of such functions is the same as the undeformed one.  The effect
of the noncommutativity of the plane is only felt when $x^+$ and $x^-$
are put together. Consider the remaining commutator $
\left[J_+(x),J_-(y)\right]_\star $. For this we have to use
\begin{equation}
e^{\frac{i}{2} \del_x\wedge\del_y} e^{- ikx^+ } e^{ ipy^-}
    = e^{-i\theta k p} e^{-ikx^+ } e^{ ipy^-}\ .
\label{twprod}
\end{equation}
It is not difficult to see that the commutation
relations~\eqn{standcomma} do not give two commuting currents, and
the theory would not be conformally invariant in any sense. It is
however possible to still obtain two mutually commuting algebras
with the following \emph{deformation of  the commutation relations
}\eqn{standcomma}:
\begin{equation}
a_\sigma (p) a_{\sigma'}(q) = \mathcal{F}^{-1}(p,q) a_{\sigma'}(q)
a_\sigma (p) + 2p \delta(p+q)\delta_{\sigma\sigma'},\label{acomm}
\end{equation}
where
\begin{equation}
\mathcal{F}^{-1}(p,q)=e^{-\frac{i}{2} p\wedge q}=e^{-
i\theta(|p| q - |q| p)}\
\end{equation}
is the inverse of the twist~\eqn{twist} in momentum space. Notice
that the action of $\mathcal{F}^{-1}(p,q)$ is always zero when
considering the commutator between currents of the same chirality
because $p$ and $q$ have the same sign.  Summarizing, thanks to
relations~\eqn{acomm} left and right currents commute, while
currents of the same chirality yield a central term, as in the
standard theory. This result, whose main ingredient is Eq.
\eqn{acomm}, descends from the choice of the  star-commutator  in
the form
\be
[ \phi(x),  \phi(y)]_\star\equiv \phi(x)\star  \phi(y)-
\phi(y)\star \phi(x).
\ee
Indeed, \eqn{acomm} imply in particular that the star-commutator
of the $\phi$ fields is the standard one. Other choices could be
undertaken (see for example \cite{bu, zhan}), all of which equally
legitimate, in the absence of a guiding principle, which is for
the moment still missing.

The appearance in \eqn{acomm} of a ``quantum plane-like''
deformation of the commutation between creation and annihilation
operators in a quantum field theory on the noncommutative plane is
not new. It has already appeared in higher dimensional
theories~\cite{bmpv} where it is postulated a relation of the kind
$a(p)a(q)=G(p,q)a(q)a(p)$ to preserve Lorentz invariance. The use of
the twisted coproduct (and the proper limit for $\theta\to 0$) then
fixes the quantity $G$ to be the twist in dimensions greater than 2.
In $1+1$ dimensions Lorentz invariance is not enough to fix the
precise form of the deformation of the commutation relations.

Now that we have the Ka\c{c}-Moody currents, we demonstrate our
initial conjecture, that is, we construct the stress-energy tensor in
terms of the Ka\c{c}-Moody currents and verify that its components
generate the standard quantum conformal algebra.

Using the relations~\eqn{acomm} and the twist, one can see that
the commutators $[\Theta^\pm(x),\Theta^\pm(y)]$ are unchanged from
the usual case (see for example \cite{FHJ}), as the twist has no
effect on pairs of functions of $x^+$ (or $x^-$) only:
\begin{equation}
[\Theta^\pm(x),\Theta^\pm(y)]_\star = \pm \frac{4 i}{\pi} \Theta^{\pm
\pm}(x)\del_{\mp} \delta(x-y) - \frac{i}{6 \pi^3} \del_{\mp}^{'''}
\delta(x-y)\ .
\end{equation}
The mixed commutator instead implies a nontrivial combination of
\eqn{twprod}, \eqn{acomm}.  A straightforward, although tedious,
calculation gives
\begin{equation}
[\Theta^{++}(x),\Theta^{--}(y)]_\star =0\ ,
\end{equation}
thus confirming the existence of twisted conformal symmetry.

In this article we have shown that it is possible to build a
2-dimensional quantum field theory on the noncommutative plane which
still possesses an infinite dimensional symmetry. This symmetry is a
quantum deformation of the standard conformal algebra with
undeformed Lie brackets. In~\cite{BLRV06} and~\cite{alvarez} an
interpretation has been suggested of twisted symmetries. While
\cite{alvarez} argues that these are not standard physical
symmetries because they modify the star product, in \cite{BLRV06},
upon performing an analysis of covariance of noncommutative field
theories, it is shown that, while covariance amounts to invariance
under ``observer'' dependent transformations, twist symmetries are
``particle'' dependent transformations, namely, transformations
which modify localized fields but not background fields as the
deformation tensor $\theta$ (for details on ``observer'' and
``particle'' dependent transformations we refer to
\cite{Colladay:1998fq}).

 The task would now be to use this conformal
theory to unveil in the noncommutative case as well, all the richness
of this infinite quantum symmetry, exploring for example possible
novel relations among vertex operators which come from the twist.
We plan to return on this aspect, as well as a more detailed
description of the symmetry, in a longer report in the near
future.

It is a pleasure to thank P.~Aschieri, A.~P.~Balachandran,
S.~Kurkcuoglu and J.~Lukierski for discussions and
correspondence. This work has been supported in part by the {\sl
Progetto di Ricerca di Interesse Nazionale {\em Sintesi}}.

\end{document}